\documentclass[prd,aps,floats,preprintnumbers,preprint]{revtex4}

\usepackage{graphicx}
 
\newcommand{\be}{\begin{equation}}
\newcommand{\ee}{\end{equation}}
\newcommand{\bea}{\begin{eqnarray}}
\newcommand{\eea}{\end{eqnarray}}

\begin{document}

%\preprint{YITP-09-35}

\setlength{\unitlength}{1mm}

\title{Spatial averaging and apparent acceleration in inhomogeneous spaces}

\author{Antonio Enea Romano$^{1,2,3}$ and Misao Sasaki$^1$}
\affiliation{
$^1$Yukawa Institute for Theoretical Physics, Kyoto University, Kyoto 606-8502, Japan\\
$^2$Leung Center for Cosmology and Particle Astrophysics, National Taiwan University, Taipei 10617, Taiwan, R.O.C.\\
$^3$Instituto de Fisica, Universidad de Antioquia, A.A.1226, Medellin, Colombia\\
}

%\author{Antonio Enea Romano$^{1,2,3}$ \and Misao Sasaki$^1$}
%\institute{Antonio Enea Romano \and Misao Sasaki
%\at Yukawa Institute for Theoretical Physics, Kyoto University, Kyoto 606-8502, Japan\\
%$^2$Leung Center for Cosmology and Particle Astrophysics, National Taiwan University, Taipei 10617, Taiwan, R.O.C.\\
%$^3$Instituto de Fisica, Universidad de Antioquia, A.A.1226, Medellin, Colombia\\
%}

\begin{abstract}
As an alternative to dark energy that explains the observed acceleration
of the universe, it has been suggested that we may be at the center
 of an inhomogeneous isotropic universe described by
 a Lemaitre-Tolman-Bondi (LTB) solution of Einstein's field equations.
To test this possibility, it is necessary to solve the null geodesics.
In this paper we first give a detailed derivation of a fully analytical set of differential equations for 
the radial null geodesics as functions of the redshift in LTB models.
As an application we use these equaions to show that a positive averaged acceleration $a_D$ 
obtained in LTB models through spatial averaging can be incompatible
 with cosmological observations.
We provide examples of LTB models with positive $a_D$ which fail to
 reproduce the observed luminosity distance $D_L(z)$. 
Since the apparent cosmic acceleration $a^{FLRW}$  is obtained from fitting the 
observed luminosity distance to a FLRW model we conclude that 
in general a positive $a_D$ in LTB models does not imply a positive $a^{FLRW}$. 
\end{abstract}
\maketitle

\section{Introduction}
High redshift luminosity distance measurements \cite{Perlmutter:1998np,
Riess:1998cb,Tonry:2003zg,Knop:2003iy,Barris:2003dq,Riess:2004nr} and
the WMAP measurement \cite{WMAP2003,Spergel:2006hy} of cosmic
microwave background (CMB) interpreted in the context of standard 
FLRW cosmological models have strongly disfavored a matter dominated universe,
 and strongly supported a dominant dark energy component, giving rise
to a positive cosmological acceleration, which we will denote by $a^{FLRW}$
 (not to be confused with the scale factor $a$). As an alternative to dark
energy, it has been proposed \cite{Nambu:2005zn,Kai:2006ws}
 that we may be at the center of an inhomogeneous isotropic universe 
described by a Lemaitre-Tolman-Bondi (LTB) solution of Einstein's field 
equations, where spatial averaging over one expanding and one contracting 
region is producing a positive averaged acceleration $a_D$.
Another more general approach is to directly map luminosity distance
 as a function of redshift $D_L(z)$ to LTB 
models \cite{Celerier:1999hp,Iguchi:2001sq},
and more recently different groups \cite{Chung:2006xh,Yoo:2008su}
 have shown that an inversion method can be applied successfully to 
reproduce the observed $D_L(z)$.
The main point is that the luminosity distance is in general sensitive 
to the geometry of the space through which photons are propagating along
null geodesics, and therefore arranging appropriately the geometry of 
a given cosmological model it is possible to reproduce a given $D_L(z)$.

The averaged acceleration $a_D$ on the other side is not directly 
related to $a^{FLRW}$, since the latter is obtained by integrating 
the position dependent cosmological redshift along the null geodesics, 
while $a_D$ is the result of spatial averaging, and has no relation to 
the causal structure of the underlying space.
It was shown \cite{Romano:2006yc} that this crucial difference 
can make $a_D$ unobservable to a central observer $O_c$ when the
 scale of the spatial averaging is greater than its event horizon.

In this paper we will further investigate the relation between LTB models
 with positive averaged acceleration and cosmological observations, 
showing that in general they can be incompatible.
Different authors have studied acceleration in LTB 
spaces \cite{Hirata:2005ei,Vanderveld:2006rb} using different definitions,
showing that the deceleration parameter $q_0$ deduced from the
luminosity distance observation cannot be negative if the origin
of an LTB model is regular and smooth. But they did not focus on 
models with spatially averaged acceleration as we do.

It should also be mentioned the work by \cite{Bolejko:2008yj}, where
some models with positive averaged acceleration are shown to be unrealistic.
On the other hand, what we show is that for models with positive 
averaged acceleration, $q(z)$ apparent defined from the luminosity
distance $D_L(z)$ is not negative, which in principle
could have become negative, independent of whether
such models are unrealistic or not.

\section{Lemaitre-Tolman-Bondi (LTB) Solution\label{ltb}}
Lemaitre-Tolman-Bondi  solution can be
 written as \cite{Lemaitre:1933qe,Tolman:1934za,Bondi:1947av}
\begin{eqnarray}
\label{eq1} %
ds^2 = -dt^2  + \frac{\left(R,_{r}\right)^2 dr^2}{1 + 2\,E}+R^2
d\Omega^2 \, ,
\end{eqnarray}
where $R$ is a function of the time coordinate $t$ and the radial
coordinate $r$, $R=R(t,r)$, $E$ is an arbitrary function of $r$, $E=E(r)$
and $R,_{r}=\partial R/\partial r$.

Einstein's equations give
\begin{eqnarray}
\label{eq2} \left({\frac{\dot{R}}{R}}\right)^2&=&\frac{2
E(r)}{R^2}+\frac{2M(r)}{R^3} \, , \\
\label{eq3} \rho(t,r)&=&\frac{2 M,_{r}}{R^2 R,_{r}} \, ,
\end{eqnarray}
with $M=M(r)$ being an arbitrary function of $r$ and the dot denoting
the partial derivative with respect to $t$, $\dot{R}=\partial R(t,r)/\partial t$.
 The solution of Eq.\ (\ref{eq2}) can be expressed parametrically 
in terms of a time variable $\eta=\int^t dt'/R(t',r) \,$ as
\begin{eqnarray}
\label{eq4} \tilde{R}(\eta ,r) &=& \frac{M(r)}{- 2 E(r)}
     \left[ 1 - \cos \left(\sqrt{-2 E(r)} \eta \right) \right] \, ,\\
\label{eq5} t(\eta ,r) &=& \frac{M(r)}{- 2 E(r)}
     \left[ \eta -\frac{1}{\sqrt{-2 E(r)} } \sin \left(\sqrt{-2 E(r)}
     \eta \right) \right] + t_{b}(r) \, ,
\end{eqnarray}
where  $\tilde{R}$ has been introduced to make clear the distinction
 between the two functions $R(t,r)$ and $\tilde{R}(\eta,r)$
 which are trivially related by 
\begin{equation}
R(t,r)=\tilde{R}(\eta(t,r),r) \, ,
\label{Rtilde}
\end{equation}
and $t_{b}(r)$ is another arbitrary function of $r$, called the bang function,
which corresponds to the fact that big-bang/crunches can happen at different
times. This inhomogeneity of the location of the singularities is one of
the origins of the possible causal separation \cite{Romano:2006yc} between 
the central observer and the spatially averaged region for models
 with positive $a_D$.

We introduce the variables
\begin{equation}
 a(t,r)=\frac{R(t,r)}{r},\quad k(r)=-\frac{2E(r)}{r^2},\quad
  \rho_0(r)=\frac{6M(r)}{r^3} \, ,
\end{equation}
so that  Eq.\ (\ref{eq1}) and the Einstein equations
(\ref{eq2}) and (\ref{eq3}) are written in a form 
similar to those for FLRW models,
\begin{equation}
\label{eq6} ds^2 =
-dt^2+a^2\left[\left(1+\frac{a,_{r}r}{a}\right)^2
    \frac{dr^2}{1-k(r)r^2}+r^2d\Omega_2^2\right] \, ,
\end{equation}
\begin{eqnarray}
\label{eq7} %
\left(\frac{\dot{a}}{a}\right)^2 &=&
-\frac{k(r)}{a^2}+\frac{\rho_0(r)}{3a^3} \, ,\\
\label{eq:LTB rho 2} %
\rho(t,r) &=& \frac{(\rho_0 r^3)_{, r}}{3 a^2 r^2 (ar)_{, r}} \, .
\end{eqnarray}
The solution of Eqs.\ (\ref{eq4}) and (\ref{eq5}) can now be written as
\begin{eqnarray}
\label{LTB soln2 R} \tilde{a}(\tilde{\eta},r) &=& \frac{\rho_0(r)}{6k(r)}
     \left[ 1 - \cos \left( \sqrt{k(r)} \, \tilde{\eta} \right) \right] \, ,\\
\label{LTB soln2 t} t(\tilde{\eta},r) &=& \frac{\rho_0(r)}{6k(r)}
     \left[ \tilde{\eta} -\frac{1}{\sqrt{k(r)}} \sin
     \left(\sqrt{k(r)} \, \tilde{\eta} \right) \right] + t_{b}(r) \, ,
\end{eqnarray}
where $\tilde{\eta} \equiv \eta\, r = \int^t dt'/a(t',r) \,$.

In the rest of paper we will use this last set of equations 
and drop the tilde to make the notation simpler.
Furthermore, without loss of generality, we may set 
the function $\rho_0(r)$ to be a constant,
 $\rho_0(r)=\rho_0=\mbox{constant}$.

\section {Geodesic equations}
The luminosity distance for a central observer in a LTB space 
as a function of the redshift is expressed as
\be
D_L(z)=(1+z)^2 R\left(t(z),r(z)\right)
=(1+z)^2 r(z)a\left(\eta(z),r(z)\right) \,,
\ee
where $\Bigl(t(z),r(z)\Bigr)$ or $\Bigl((\eta(z),r(z)\Bigr)$
is the solution of the radial geodesic equation
as a function of the redshift.
The past-directed radial null geodesic is given by
\bea
\label{geo1}
\frac{dT(r)}{dr}=f(T(r),r) \,,
\quad
f(t,r)=\frac{-R_{,r}(t,r)}{\sqrt{1+2E(r)}} \,,
\eea
where $T(r)$ is the time coordinate along the null radial geodesic as a function of the the coordinate $r$.
Applying the definition of red-shift it is possible to obtain \cite{Celerier:1999hp}:

\begin{eqnarray}
{dr\over dz}&=&{\sqrt{1+2E(r(z))}\over {(1+z)\dot {R'}[T(r(z)),r(z)]}} \,,
\label{eq:34} \\
{dt\over dz}&=&-\,{R'[T(r(z),r(z))]\over {(1+z)\dot {R'}[T(r(z)),r(z)]}} . 
\label{eq:35} \\
\end{eqnarray}

or in terms of the function $f(t,r)$
\bea
{dr\over dz}&=&-{1\over{(1+z)\dot{f}(t(z),r(z))}}\,,\label{rz} \\
{dt\over dz}&=&-{f(t(z),r(z))\over{(1+z)\dot{f}(t(z),r(z))}}\,. \label{tz}
\eea

In order to solve the above differential equations we need $R(t,r)$ which can only be obtained numerically by integrating the Einstein's equations, while the analytical solution is expressed in terms of $a(\eta,r)$. For this reason it is convenient to re-write the above equations in terms of the coordinates $(\eta,r)$.
From the implicit solution, we can write 
\bea
T(r)=t(U(r),r)\,, \\
\frac{dT(r)}{dr}=\frac{\partial t}{\partial \eta} \frac{dU(r)}{dr}+\frac{\partial t}{\partial r} \,,
\eea
where $U(r)$ is the $\eta$ coordinate along the null radial geodesic as a function of the the coordinate $r$.
In order to perform the change of variables from $(t,r)$ to $(\eta,r)$ we need to use the fact that the derivation of the implicit  solution $a(\eta,r)$ is based on the use of 
the conformal time variable $\eta$, which by construction 
satisfies the relation
\be
\frac{\partial\eta(t,r)}{\partial t}=a^{-1} \,,
\ee
from which, after a careful treatment of the partial derivatives, we can derive the following relations:
\bea
t(\eta,r)&=&t_b(r)+\int^{\eta}_{0}a(\eta^{'},r) d\eta^{'} \, ,\\
dt&=&a(\eta,r)d\eta+\left(\int^{\eta}_{0} \frac{\partial a(\eta^{'},r)}{\partial r} d\eta^{'}+t_b^{'}(r)\right) dr \,,\\
%&&t(0,r)=t_b(r)\,, \quad a(0,r)=0 \, , \\
\frac{\partial}{\partial t}&=&a^{-1}{\frac{\partial}{\partial \eta}}\,, \\
%&&\frac{\partial t}{\partial r}(\eta,r)=\int^{\eta}_{0}\frac{\partial a(\eta^{'},r)}{\partial r} %d\eta^{'}+t_b^{'}(r) \, , 
\partial_r \eta&=&-a(\eta,r)^{-1}\partial_r t  \,. 
%&&\frac{\partial^2R}{\partial t\partial r}
%=a^{-1}\left(\frac{\partial^2 R(\eta,r)}{\partial \eta \partial r}
%+\frac{\partial^2 R(\eta,r)}{\partial \eta^2}\frac{\partial \eta}{\partial r}
%+\frac{\partial R(\eta,r)}{\partial \eta}\frac{\partial}{\partial \eta}
%\left(\frac{\partial \eta}{\partial r}\right)\right) \,.
\eea
We can now write :
\bea
{dt\over dz}={\partial t \over\partial\eta}{d\eta\over dz}+{\partial t \over\partial r}{dr \over dz}=a{d\eta \over dz}+\partial_r t{dr \over dz}
\eea 
from which using the geodesic equations (\ref{rz},\ref{tz}) we get:
\bea
{d\eta\over dz}={1\over a} {{\partial_r t-f}\over{(1+z)\dot{f}}}.
\eea
We can then express $f$ and $\dot{f}$ in terms of the analytical solution using:

\bea
f(t(\eta,r),r)&=&F(\eta,r) \,,\\
\dot{f}(t(\eta,r),r)&=&{1\over a}\partial_{\eta} F(\eta,r) \,,\\
R_{,r}(t,r)&=&\partial_r R(t(\eta,r),r)+\partial_{\eta} R(t(\eta,r),r) \partial_r \eta \,,\\
R(t(\eta,r),r)&=&r\, a(\eta,r) \,, \\
F(\eta,r)&=&-\frac{1}{\sqrt{1-k(r)r^2}}\left[\partial_r (a(\eta,r) r)
+\partial_{\eta} (a(\eta,r) r) \partial_r \eta\right]  \, \nonumber \\
&=&-\frac{1}{\sqrt{1-k(r)r^2}}\left[\partial_r (a(\eta,r) r)
-\partial_{\eta} (a(\eta,r) r) a(\eta,r)^{-1}\partial_r t \right]\,.
\eea
Finally we get:
\bea
\label{geo3}
\frac{d \eta}{dz}
&=&\frac{\partial_r t(\eta,r)-F(\eta,r)}{(1+z)\partial_{\eta}F(\eta,r)}=p(\eta,r) \,,\\
\label{geo4}
\frac{dr}{dz}
&=&-\frac{a(\eta,r)}{(1+z)\partial_{\eta}F(\eta,r)}=q(\eta,r) \,. 
\eea
where $\eta=U(r(z))$ is the value of the variable $\eta$ along the null geodesic.

It is important to observe that the functions $p,q,F$ have an explicit analytical form which can be obtained from $a(\eta,r)$ and $t(\eta,r)$ using the relations above.
In this way the coefficients of equations (\ref{geo3}) and (\ref{geo4}) are 
fully analytical, which is a significant improvement over previous 
approaches which required a numerical integration of the Einstein's equations.
This version of the geodesics equations takes full advantage of the existence of the analytical solution and is suitable for both numerical and analytical applications such as low red-shift expansions.

\section{Averaged acceleration }
Following the standard averaging 
procedure \cite{Buchert:1999mc,Palle:2002zf,Nambu:2005zn}
we define the volume for a spherical domain, $0<r<r_D$, as 
\begin{equation} \label{eq:ltb volume}
V_D %
= 4 \pi \int_0^{r_D} \frac{R^2 R,_{r}}{\sqrt{1+2E(r)}} \, dr \, ,
\end{equation}
and the length associated with the domain as
\begin{equation}
\label{LD}
L_D = V_D^{1/3}.
\end{equation}
Then the deceleration parameter $q_D$ and the averaged
 acceleration $a_D$ (not to be confused with the scale factor $a$) 
are defined as
\begin{eqnarray}
\label{qD}
q_D & = & -\ddot{L}_D L_D/\dot{L}_D^2 \, ,\\
\label{aD}
a_D & = & \ddot{L}_D/L_D\, .
\end{eqnarray}

As models that give a positive averaged acceleration,
we consider those studied in \cite{Chuang:2005yi}.
They are characterized by the functions $k(r)$ and $t_b(r)$
given by
\begin{eqnarray}
t_b(r) &=& - \frac{h_{tb}(r/r_t)^{n_t}}{1+(r/r_t)^{n_t}}    \, , 
\\
k(r) &=& -\frac{(h_k+1) (r/r_k)^{n_k}}{1+(r/r_k)^{n_k}}+1 \, .
\end{eqnarray}
Note that they fix the length scale by setting $k(0)=1$.
After exploring the 9 parameters space
 they give three examples 
of LTB solutions with positive $q_D$ as shown in Table I.

\tabcolsep=8pt
\begin{table}[ht!]
\caption{Three examples of the domain acceleration.} \label{table:domain eg} %
\center
\begin{tabular}{|c|c|c|c|c|c|c|c|c|c||c|}\hline
&$t$ & $r_D$ & $\rho_0$ & $r_k$ & $n_k$ & $h_k$ & $r_t$ & $n_t$ & $h_{tb}$ & $q_D$
\\
\hline
1&$0.1$ & $1$ & $1$ & $0.6$ & $20$ & $10$ & $0.6$ & $20$ & $10$ & $-0.0108$ 
\\
\hline
2&$0.1$ & $1.1$ & $10^5$ & $0.9$ & $40$ & $40$ & $0.9$ & $40$ & $10$ & $-1.08$ 
\\
\hline
3&$10^{-8}$ & $1$ & $10^{10}$ & $0.77$ & $100$ & $100$ & $0.92$ & $100$ & $50$
 & $-6.35$ 
\\
\hline
\end{tabular}

\begin{tabular}{|c|c|c|c|c|}
\hline
& $L_D$ & $\dot{L}_D$ & $\ddot{L}_D$ & $q_D$\\
\hline
1&$16.2$&$1.62$&$0.00174$& $-0.0108$\\
\hline
2& $94.0$ & $7.63$ & $0.694$ & $-1.08$ \\
\hline
3&$8720$&$117$&$10.0$& $-6.35$\\
\hline
\end{tabular}
\end{table}

Defining $t_q$ as the time in the first column of Table I, i.e. the 
time at which $q(t_q)=q_D$, with $q_D$ being the value in the last 
column of the same table, we solved the null geodesics
 equations (\ref{geo3}) and (\ref{geo4}) imposing the following 
initial conditions.
\begin{eqnarray}
\label{INI1}
\eta(z=0) & = &\eta_q=\eta(t_q,0) \, ,  \\
\label{INI2}
 r(z=0)& = & 0 \, ,
\end{eqnarray}
where $\eta(t_q,0)$ is obtained by solving numerically
 for $\eta_q$, Eq.~(\ref{LTB soln2 t}),
\be
t(\eta_q,0)=t_q \, ,
\ee
For the second and third model of Table I which we will 
study we obtain, respectively,
\bea
\eta_q&=&0.0330199 \,,  \\
\eta_q&=&3.30189\times 10^{-6} \,.
\eea
Eq.~({\ref{INI1}}) is the natural way to map these models into 
luminosity distance observations for a central observer which should 
receive the light rays at the time $t_q$ at which the averaged 
acceleration is positive. 

\section{Comparing LTB to observations}
An important model independent quantity is the expansion history, $H(z)$,
whose value can be reconstructed from observations of the luminosity 
distance, $D_L$,
via a single differentiation
\cite{st98,HT99,chiba99,saini00,chiba00}
\be\label{eq:H}
H(z)=\left[{d\over dz}\left({D_L(z)\over 1+z}\right)\right]^{-1}~.
\ee

\begin{figure}[h]
\begin{center}
\includegraphics[height=60mm,width=80mm]{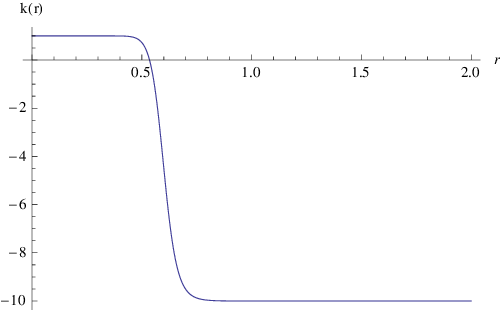}
\includegraphics[height=60mm,width=80mm]{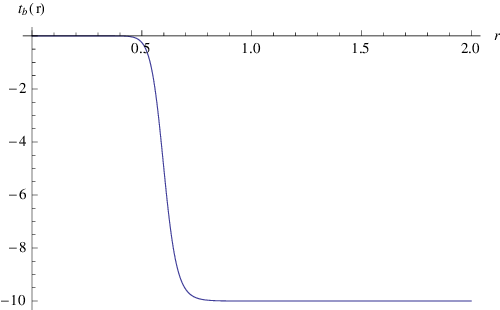}
\caption{$t_b(r)$ and $k(r)$ are plotted for the second model in Table I.}
\end{center}
\end{figure}

\begin{figure}[h]
\begin{center}
\includegraphics[height=60mm,width=80mm]{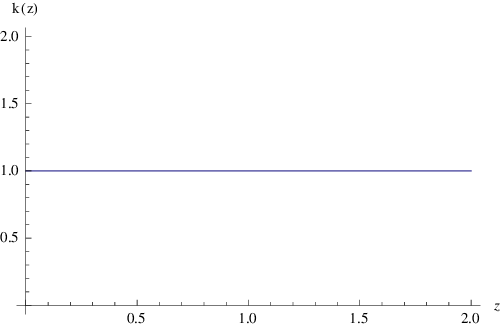}
\includegraphics[height=60mm,width=80mm]{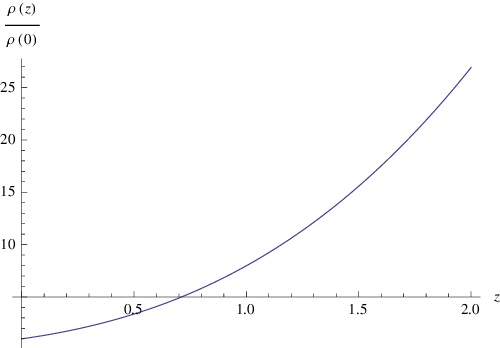}
\caption{$k(r(z))$ and $\rho(\eta(z),r(z))$ are plotted
 for the second model in Table I.}
\end{center}
\end{figure}

\begin{figure}[h]
\begin{center}
\includegraphics[height=60mm,width=80mm]{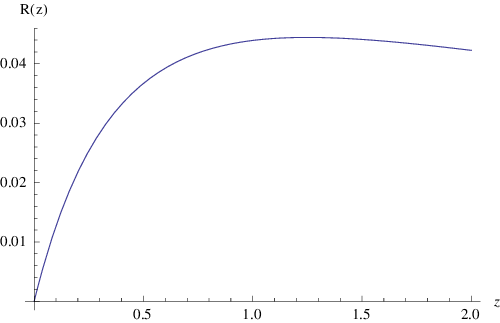}
\includegraphics[height=60mm,width=80mm]{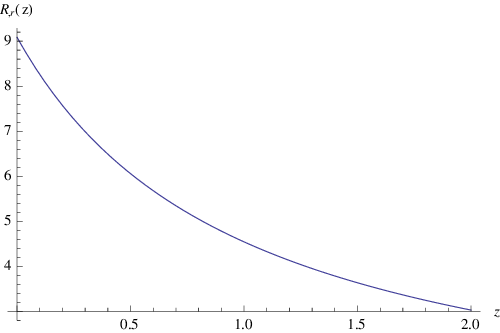}
\caption{ $R(\eta(z),r(z))$ and $R_{,r}(\eta(z),r(z))$ are 
plotted for the second model in Table I.}
\end{center}
\end{figure}

\begin{figure}[h]
\begin{center}
\includegraphics[height=60mm,width=80mm]{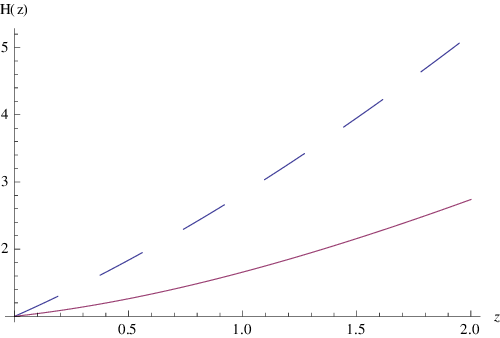}
\includegraphics[height=60mm,width=80mm]{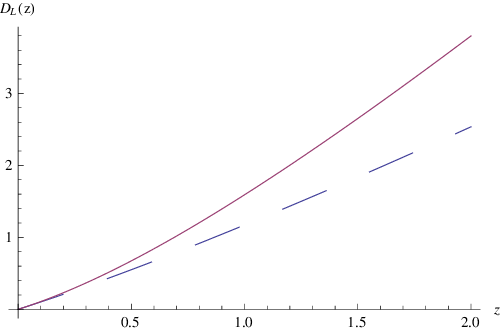}
\caption{ $H(z)$ and $D_L(z)$ are plotted for the second model in Table I.
The dashed line corresponds to the concordance $\Lambda$CDM model.
}
\end{center}
\end{figure}

\begin{figure}[h]
\begin{center}
\includegraphics[height=60mm,width=80mm]{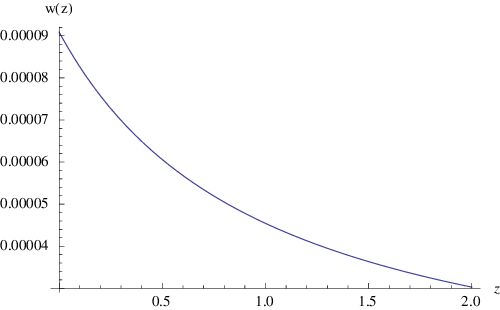}
\includegraphics[height=60mm,width=80mm]{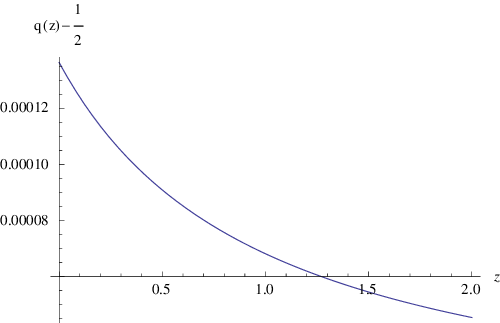}
\caption{ $w(z)$ and $q(z)$ are plotted for the second model in Table I.}
\end{center}
\end{figure}

\begin{figure}[h]
\begin{center}
\includegraphics[height=60mm,width=80mm]{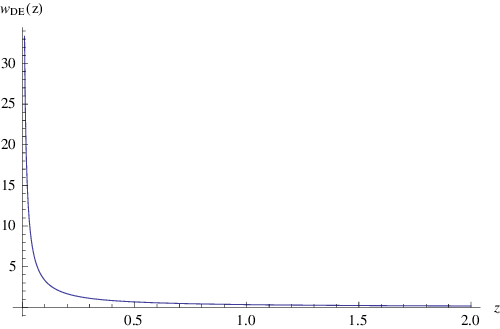}
\caption{ $w_{DE}(z)$ is plotted for the second model in Table I.}
\end{center}
\end{figure}

\begin{figure}[h]
\begin{center}
\includegraphics[height=60mm,width=80mm]{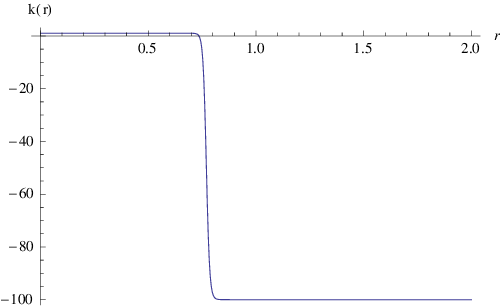}
\includegraphics[height=60mm,width=80mm]{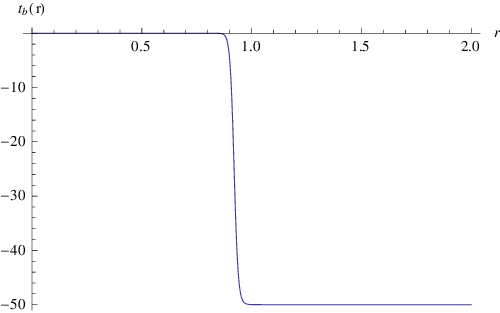}
\caption{ $t_b(r)$ and $k(r)$ are plotted for the third model in Table I.}
\end{center}
\end{figure}

\begin{figure}[h]
\begin{center}
\includegraphics[height=60mm,width=80mm]{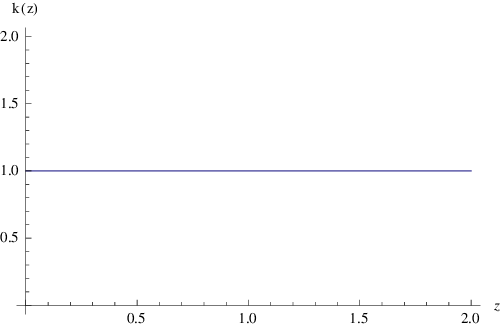}
\includegraphics[height=60mm,width=80mm]{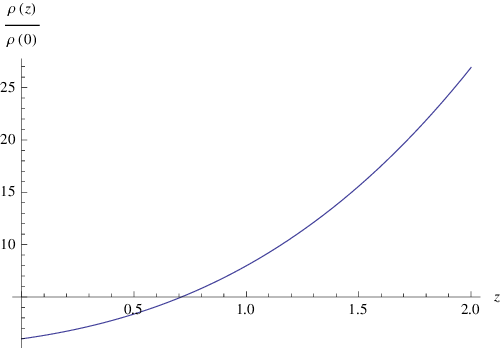}
\caption{ $k(r(z))$ and $\rho(\eta(z),r(z))$ are plotted for the third 
model in Table I.}
\end{center}
\end{figure}

\begin{figure}[h]
\begin{center}
\includegraphics[height=60mm,width=80mm]{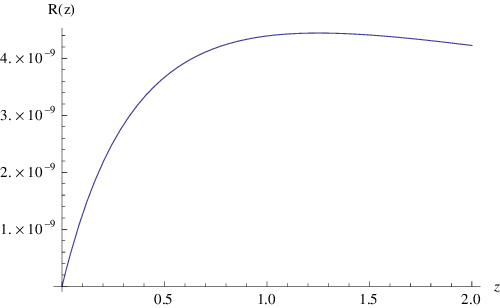}
\includegraphics[height=60mm,width=80mm]{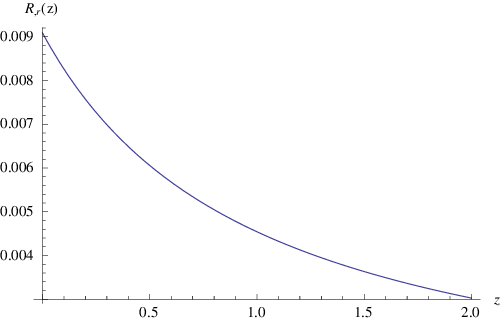}
\caption{ $R(\eta(z),r(z))$ and $R_{,r}(\eta(z),r(z))$ are plotted 
for the third model in Table I.}
\end{center}
\end{figure}

\begin{figure}[h]
\begin{center}
\includegraphics[height=60mm,width=80mm]{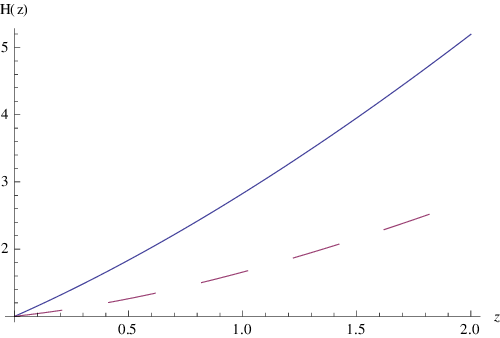}
\includegraphics[height=60mm,width=80mm]{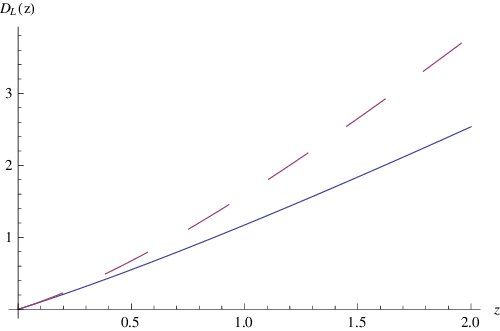}
\caption{ $H(z)$ and $D_L(z)$ are plotted for the third model in Table I.
 The dashed line corresponds to the concordance $\Lambda$CDM model.}
\end{center}
\end{figure}

\begin{figure}[h]
\begin{center}
\includegraphics[height=60mm,width=80mm]{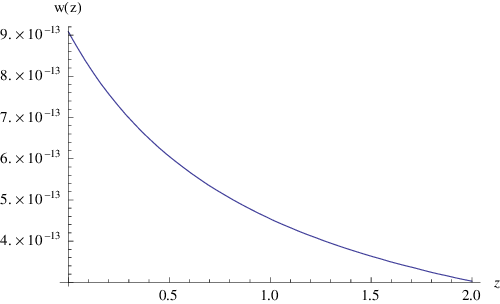}
\includegraphics[height=60mm,width=80mm]{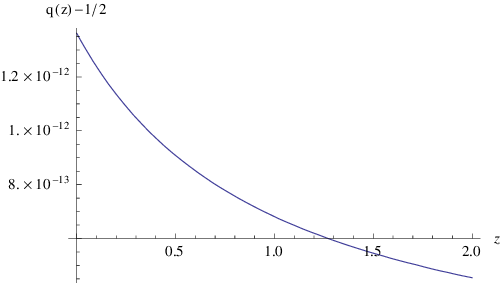}
\caption{ $w(z)$ and $q(z)$ are plotted for the third model in Table I.}
\end{center}
\end{figure}

\begin{figure}[h]
\begin{center}
\includegraphics[height=60mm,width=80mm]{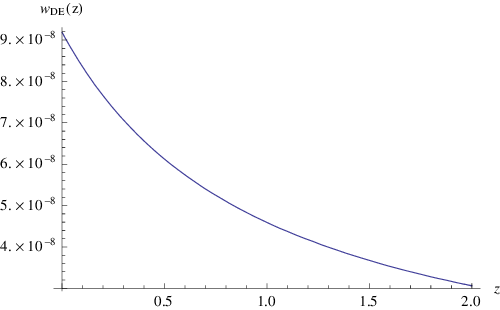}
\caption{ $w_{DE}(z)$ is plotted for the third model in Table I.}
\end{center}
\end{figure}

Note that this assumes the universe is not only homogeneous and isotropic
but also spatially flat.
The equation of state, $w(z)$, of DE is more cumbersome to reconstruct since it
involves second derivatives of $D_L(z)$ and is therefore a noisier quantity
than $H(z)$. An additional source of uncertainty relating to $w(z)$
is caused by the fact that the value of the matter density, $\Omega_{0m}$ enters
into the determination of $w(z)$ explicitly, through the expression
\be\label{eq:state}
w_{DE}(x) =
\frac{(2 x /3) \ d \ {\rm ln}H \ / \ dx - 1}{1 \ - \ (H_0/H)^2
\Omega_{0m} \ x^3}\,\,.
\ee
In our case we can first calculate $D_L(z)$ according to the LTB
 solutions geodesics equation we derived in the previous section, 
and then use it to construct $H(z)$ and $w(z)$ according to 
Eqs.~(\ref{eq:state}) and (\ref{eq:H}).

We compare $H(z)$ and $D_L(z)$ to the observed $H_F$ and $D_L^F$ 
using the standard best fitted FLRW cosmological model,
\bea
H_{F}(z)^2=H_0^2\left(\Omega_m(1+z)^3+(1-\Omega_m)\right) && \Omega_m=0.25 \, , \\
D_L^F(z)=(1+z)\int^z_0\frac{1}{H_{F}(x)} dx \, ,
\eea
It should be noted that the parameters given in Table I require to fix the units,
and we do that by imposing
\be
H(0)=H_0 \,.
\ee
In this way we eliminate any ambiguity in the choice of units
in order to be able to compare the observed $H_F(z)$ and $H(z)$.

For both the second and third models in Table I, $H(z)$ and $D_L(z)$ 
are not reproducing correctly observational data both from the quantitative 
and qualitative points of view. The effective $w(z)$ is small and positive 
and the effective $w_{DE}(z)$ is not $-1$ for both the second and third models.
On the contrary, for the redshift range of $z \lesssim 1$, 
$w_{DE}(z)> 1$ for the second model, while $1\gg w_{DE}(z)>0$ 
for the third model.

As it can be seen from the plots, $k(z)$ is approximatively 
constant in the observationally interesting redshift range,
but given the parameters of these models matter is 
dominating at the time $t_q$, since as shown above $\eta_q\ll1$,
which explains why $w$ is so small and $q(z)\approx 1/2$.

These examples show how a positive $a_D$ does not imply a luminosity
 distance $D_L(z)$ compatible with observations, and gives a reverse 
example of the results obtained in \cite{Enqvist:2006cg}, 
where they obtained a LTB model which fits the observed luminosity 
distance, consistent with a positive $a^{FLRW}$, but without 
positive averaged acceleration $a_D$.
Our results do not rule out LTB models as alternatives to dark 
energy since  the inversion method \cite{Chung:2006xh, Yoo:2008su} 
allows us to obtain the observed luminosity distance without any averaging,
and some concrete examples derived independently have already been
 proposed \cite{Alnes:2006uk,Alnes:2005rw,Enqvist:2006cg,Alexander:2007xx,Biswas:2007gi,Tomita:2000jj}.

\section{Discussion}
We have derived a set of differential equations for the radial 
null geodesics in LTB space-time which takes full advantage of the anaytical solution.
We have then used them to compute the luminosity distance for models which have 
positive spatial acceleration and extended our analysis to other observables such
 as $H(z)$ and $w(z)$, explicitly showing that they are not 
compatible with the $\Lambda CDM$ best fit model .
Our conclusions do not rule out LTB models as alternatives to dark energy, 
but provide further evidence that physical quantities obtained via spatial 
averaging are not relevant to explain observational data.

In the future it will be interesting to find general analytical results 
in support of our conclusions and to use the equations we derived to
 provide a new method to solve the inversion problem of mapping the
 observed luminosity distance $D_L(z)$ to LTB models.

\begin{acknowledgments}
We thank A. Starobinsky and A. Notari for useful comments and discussions. 
A. E. Romano was supported by JSPS.
This work was also supported in part by JSPS Grant-in-Aid for Scientific 
Research (A) No.~21244033, and by JSPS 
Grant-in-Aid for Creative Scientific Research No.~19GS0219,
and by Monbukagaku-sho Grant-in-Aid for the global COE program,
"The Next Generation of Physics, Spun from Universality and Emergence".
\end{acknowledgments}

\end{document}